\documentclass{nature-2}
\usepackage{graphicx}
\usepackage{amsmath} 
\usepackage{dcolumn}
\usepackage{lineno} 
\usepackage[breaklinks=true,colorlinks=true,linkcolor=blue,urlcolor=blue,citecolor=blue]{hyperref}
\usepackage[style=base,font={sf,small}]{caption}
\usepackage{bm}
\title{Coherent control of plasma dynamics}
\author{Z.-H. He$^{1}$, B. Hou$^{1}$, V. Lebailly$^{2}$, J. A. Nees$^{1}$, K. Krushelnick$^{1}$, A. G. R. Thomas$^1$}

\begin{document}
\maketitle
\begin{affiliations}
\item{Center for Ultrafast Optical Science, University of Michigan, Ann Arbor, MI 48109-2099 USA}
\item{Polytech Paris-Sud - Universit\'e Paris-Sud, 91405 Orsay, France}
\end{affiliations}
\begin{abstract}
Coherent control of a system involves steering an interaction to a final coherent state by controlling the phase of an applied field. Plasmas support coherent wave structures that can be generated by intense laser fields. Here, we demonstrate the coherent control of plasma dynamics in a laser wakefield electron acceleration experiment. A genetic algorithm is implemented using a deformable mirror with the electron beam signal as feedback, which allows a heuristic search for the optimal wavefront under laser-plasma conditions that is not known \emph{a priori}. We are able to improve both the electron beam charge and angular distribution by an order of magnitude. These improvements do not simply correlate with having the `best' focal spot, since the highest quality vacuum focal spot produces a greatly inferior electron beam, but instead correspond to the particular laser phase front that steers the plasma wave to a final state with optimal accelerating fields.
\end{abstract}

\section{Introduction}
The concept of coherent control --- precise measurement or determination of a process through control of the phase of an applied oscillating field --- has been applied to many different systems, including quantum dynamics\cite{Warren_Sci_1993}, trapped atomic ions\cite{Wineland_JRNIST_1998}, chemical reactions\cite{Assion_Sci_1998}, Cooper pairs\cite{Nakamura_Nat_1999}, quantum dots\cite{Reithmaier_Nat_2004,Press_Nat_2008} and THz generation\cite{Xie_PRL_2006} to name but a few. A plasma wave is a coherent and deterministically evolving structure that can be generated by the interaction of laser light with plasma. It is therefore natural to assume that coherent control techniques may also be applied to plasma waves. Plasma waves produced by high power lasers have been studied intensively for their numerous applications, such as the production of ultrashort pulses by plasma wave compression\cite{Faure_PRL_2005,Schreiber_PRL_2010}, generation of extremely high power pulses by Raman amplification\cite{Malkin_PRL_1999,Trines_NP_2011}, 
for inertial confinement fusion ignition schemes\cite{Fisch_PRL_2002,Willingale_PRL_2011}, as well as for fundamental scientific investigations. In particular, laser wakefield acceleration of ultra-relativistic electron beams\cite{Tajima1979,Mangles_Nat_2004,Geddes_Nat_2004,Faure_Nat_2004}, has been a successful method for accelerating electrons to relativistic energies over a very short distance. In laser wakefield acceleration, an electron bunch `surfs' on the electron plasma wave generated by an intense laser and gains a large amount of energy. The accelerating electric field strength that the plasma wave can support can be many orders of magnitude higher than that of a conventional accelerator, which makes laser wakefield acceleration an exciting prospect as an advanced accelerator concept. However, although highly competitive in terms of accelerating gradient, beams from laser wakefield accelerator experiments are currently inferior to conventional accelerators in terms of other important  characteristics, such as energy spread and stability. In addition, due to constraints in laser wakefield technology, experimental demonstrations have predominantly been performed in single shot operation, far below the kHz-MHz repetition rates of conventional accelerators. 

In recent years, deformable mirror adaptive optical systems have been successfully implemented in high intensity laser experiments to increase the peak laser intensity by improving the beam focusability, especially in systems using high numerical aperture optics. The shape of the deformable mirror is generally determined in a closed loop where either a direct measurement of the wavefront is performed\cite{Bahk_OL_2004} or some nonlinear optical signal\cite{Albert_OL_2000,Planchon_OL_2006} is used as feedback in an iterative algorithm. The objective of adaptive optics has largely been optimization of the laser focal shape to a near diffraction-limited spot, thus producing the highest possible intensity. Adaptive optics can also be useful for certain focal profile shaping\cite{Zeng_JMO_1999,Agmy_OE_2005}, optimization of a laser machining process\cite{Campbell_JO_2007} or harmonic generation\cite{Bartels_Nat_2000,Yoshitomi_APB}.

In the following, we demonstrate that orders of magnitude improvement to electron beam properties from a laser wakefield accelerator operating at kHz repetition rate can be made, through the use of a genetic algorithm coupled to a deformable mirror adaptive optical system to coherently control the plasma wave formation. The electron image from a scintillator screen was processed and used in the fitness function as feedback for the genetic algorithm. Using this method, we were able to improve the beam properties significantly. This result was not simply due to an improvement in focal quality since a laser pulse with the `best' (highest intensity/lowest $M^2$) focus in vacuum produced a greatly inferior electron beam compared with a laser pulse optimized using the electron beam properties themselves. It was found that the focal spot optimized for electron beam production had pronounced intensity `wings'. Modifications to the phase front of the tightly focusing laser alter the light propagation, which experiences strong optical nonlinearities in the plasma, and therefore affect the plasma wave dynamics in a complex but deterministic manner.

\section{Results}
\subsection{Experimental setup and procedure}
The experiment was performed using the relativistic Lambda-Cubed ($\lambda^3$) laser system (see Methods). The output laser beam was reflected from a deformable mirror and focused onto a free-flowing argon gas plume to produce an electron beam by laser wakefield acceleration (see Methods) at 500 Hz. Electrons were measured using a scintillating screen imaged onto a lens-coupled CCD camera. The experimental setup is shown schematically in Fig.~\ref{setup}.

We first implemented a genetic algorithm for laser focus optimization using the second-harmonic signal generated from a beta barium borate ($\beta$-BBO) crystal (setup A in Fig.~\ref{setup}). The laser spot was optimized such that highest peak intensity is achieved when the second harmonic generation is strongest. Subsequently, we modified the fitness function to use a figure of merit (FOM, refer to equation \ref{FOM2} in Methods) from the electron scintillation data, calculating the inverse distance weighting (with power parameter n) to a single point $\mathbf{r_0}$ for all pixel intensities within an electron image. The pixel of the optimization point $\mathbf{r_0}$ was \emph{dynamically} adjusted during the genetic algorithm to concentrate all electron signal to the peak location of the charge distribution during each generation. The genetic algorithm was initialized using a `flat' mirror shape with 30 V for all actuators to allow immediate deformation in both directions. 

\subsection{Optimization of the electron spatial profile}
\begin{figure}
\centering
\includegraphics[width=160 mm]{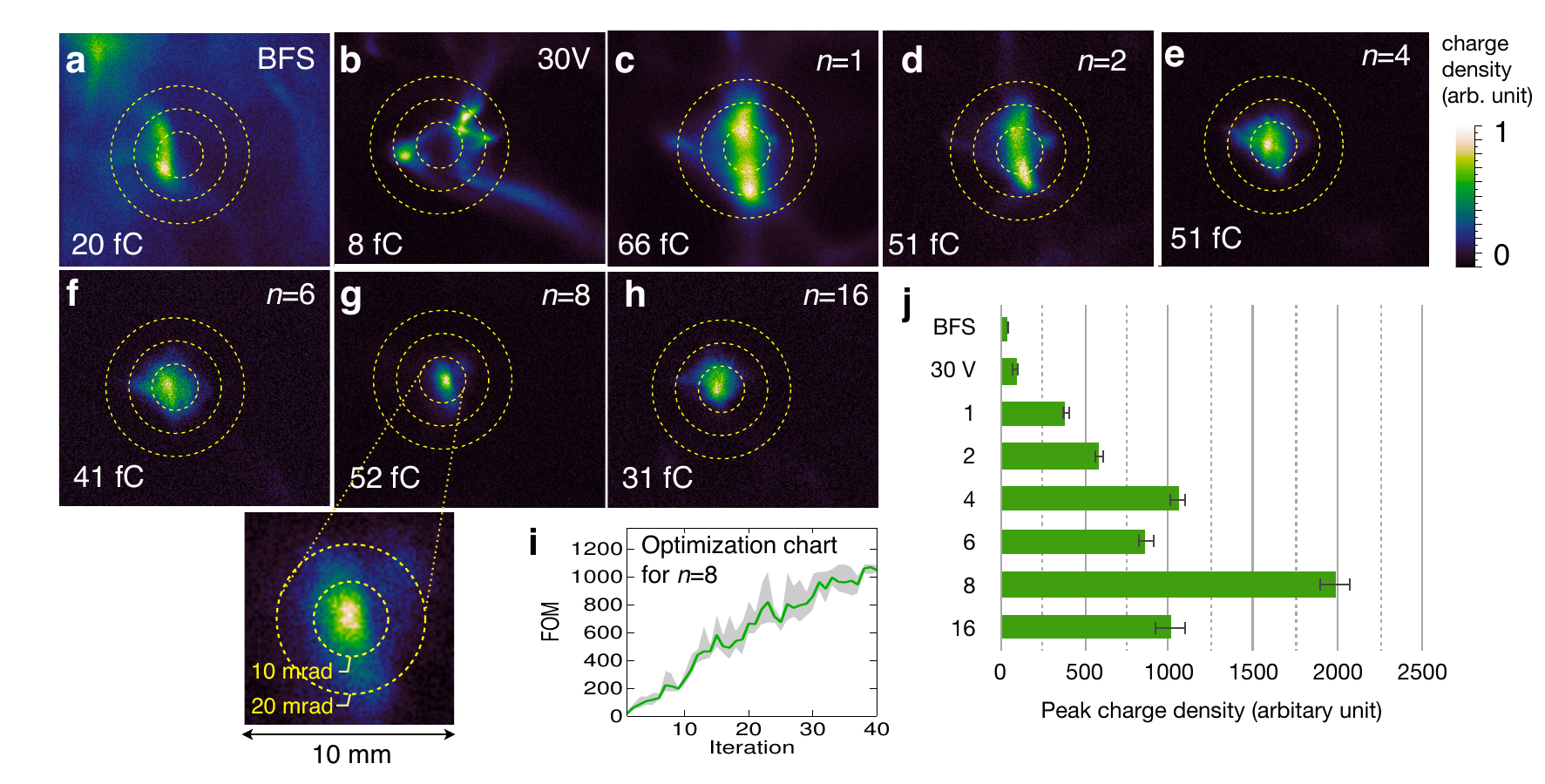}
\caption{\textbf{Optimization of the electron spatial profile.} Electron beam profile image integrated over 50 shots (100 ms exposure time) with a deformable mirror configuration (a) corrected for the best focal spot (BFS) and (b) 30 V on all actuators. (c)-(h) are single-shot electron beam profiles after genetic algorithm optimization using different weighting parameters, n. (i) shows the convergence of the genetic algorithm with n=8. The shaded gray area represents the range of the 10 best children in each iteration and the solid green curve is the average. (j) Comparison of the peak charge density in a single-shot electron image. Error bars represent the root mean square shot-to-shot fluctuations; Estimated charge per shot is displayed for every image. Contours shown are for 20, 40, 60 mrad, centered on the beam centroid.}
\label{ebeam}
\end{figure}

For comparison, electron beams produced by the `best' laser focus (by optimizing the intensity) and the initial mirror shape at 30 V are shown in Fig.~\ref{ebeam}a and b respectively. The optimized electron beam profiles are shown in Fig.~\ref{ebeam}e-j for various weighting parameters, $n$. The genetic algorithm converged to the best electron beam using $n=8$ in terms of beam divergence and peak charge density. The peak charge density was increased by a factor of 20 compared to the initial electron beam profile before optimization (see Fig.~\ref{ebeam}d). The optimized electron profile is highly stable and collimated, with a full-width-at-half-maximum (FWHM) divergence of $\Delta x=7.4\pm 0.6$ mrad and $\Delta y=12.8 \pm 1.4$ mrad. The shot-to-shot pointing (defined by the centroid position) fluctuation of the electron beam is less than 1 mrad (root mean square, r.m.s.). The integrated charge was increased by more than two-fold from the electron beams generated by a laser focus of highest intensity. The high repetition rate and real-time diagnostics permit implementation of the algorithm within a practical time frame using a standard personal computer. Typical optimization takes only a few minutes ($\sim$40 iterations) to reach convergence (see Fig.~\ref{ebeam}c).

\begin{figure}
\centering
\includegraphics[width=160 mm]{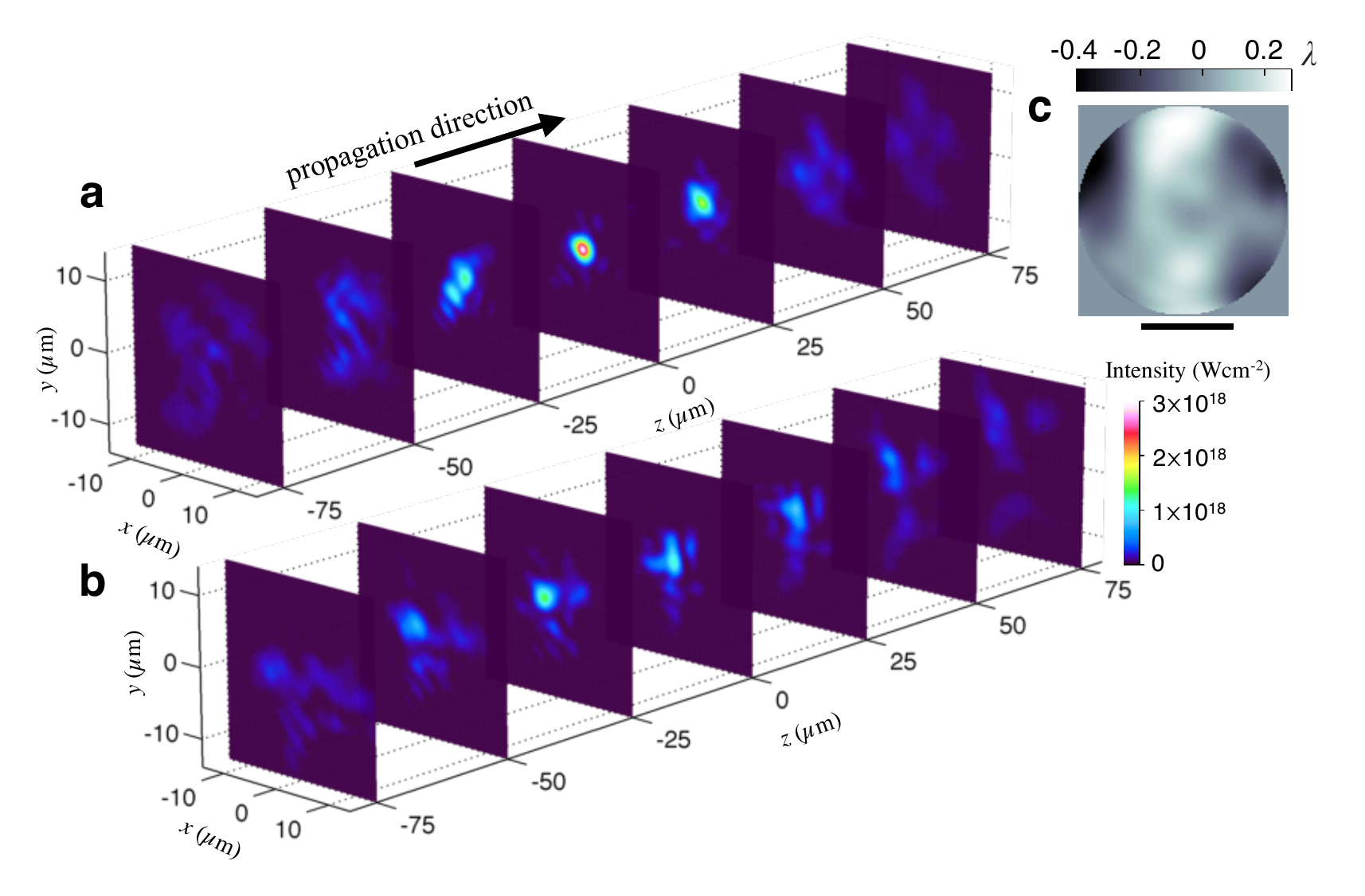}
\caption{\textbf{Laser intensity profile in the focal region and wavefront.} Scan of laser focal intensity in vacuum with the deformable mirror optimized for (a) second harmonic signal (highest intensity) and (b) electron beam in Fig.~\ref{ebeam}. (c) Relative wavefront change reconstructed from direct measurement using a Shack-Hartmann wavefront sensor. Root-mean-square phase deviation in the aperture is 0.14 wave. The wavefront was reconstructed over a slighly smaller aperture (2.26 mm diameter) than the full beam diameter on the sensor ($\approx$2.7 mm $1/e^2$ width) to reduce errors in the peripheral area. (Scale bar, 1 mm).}
\label{spot}
\end{figure}

The second harmonic optimization\cite{Albert_OL_2000} generates a near-diffraction-limited focal spot as shown in Fig.~\ref{spot}a for the far-field laser intensity profile in vacuum. In Figs.~\ref{spot}a and b, we compare the transverse intensity distribution within the focal region over the length scale of the gas jet. The laser profile (Fig.~\ref{spot}b) that produces the best electron beam exhibits several low intensity side lobes around the central peak, and has a peak intensity about half that of the optimized focus. The complex laser profiles appear to have a very dramatic effect on the structure of the plasma waves produced and consequently the electron beam profile. Fig.~\ref{spot}c shows the relative wavefront change recorded by a Shack-Hartmann wavefront sensor. Calculation using the reconstructed wavefront gives a Strehl ratio of about 0.5, which is in agreement with the far-field intensity measurement. This small wavefront modification of the driver pulse can lead to a significant improvement in the electron beam properties through the relativistic nonlinear optics of the plasma. The relative position between the focal plane and the center of the gas flow was controlled by moving the nozzle. Scanning the gas nozzle both before and after genetic algorithm confirms the optimal focal position does not change, excluding the possibility that the improvement may be due to optimizing the focal position.

\subsection{Control of energy distribution}
Furthermore, we extended the genetic algorithm optimization to \emph{control the electron energy distribution}. Through control of the light propagation, the plasma wave amplitude will be affected and therefore also the strength of the accelerating gradient. Hence, we can expect to be able to modify the energy spectrum.  A high resolution energy analyzer using a dipole magnet pair was used to obtain the electron energy spectrum as the electrons were dispersed in the horizontal plane in the magnetic field. A 150 $\mu$m pinhole was placed 2.2 cm from the electron source to improve the energy resolution of the spectrometer. The schematic setup is shown in Fig.~\ref{espec}a. The energy resolution limited by the entrance pinhole and transverse emittance of the beam is estimated to be 2 keV for the energy range of measurement.

Three rectangular masks are set in the low-, mid- and high-energy region on the dispersed data, namely masks I, II, and III in Fig.~\ref{espec}b. We employed a fitness function (see Methods) to preferentially maximize the total counts inside the mask. Raw spectra from the genetic algorithm optimization are displayed in Fig.~\ref{espec}b, showing that the brightest part has shifted congruently. The resulting spectra have mean energies of 89 keV, 95 keV and 98 keV respectively for masks I, II, and III, noting that they do not fall on the visual centroid of the image because the scintillator sensitivity is not included in the presentation of the raw data, however it was taken into account for computing the mean energies. Our results show that manipulation of electron energy distribution using the deformable mirror is somewhat restricted. The final result after optimization does not reach the objective mask completely despite that the mean energies can be varied by up to 10\%. This result is somewhat unsurprising as the scope for controlling the electron spectrum is mostly limited by the physics of the interaction - while changing the transverse intensity profile can make big differences to the shape of the plasma electric field structure, changing the maximum field amplitude of the wakefield (and therefore peak energy of the accelerated electrons) will be limited.

\subsection{Numerical simulations}
Although the details of the initial conditions required for optimal beams are difficult to determine and are therefore found using the genetic algorithm, we can at least demonstrate how modifications to the phase front of the laser pulse can improve the beam properties with an example. To illustrate the underlying physics of the plasma wave dynamics determined by the conditions of the driving laser pulse, we performed two dimensional (2D) particle-in-cell (PIC) simulations using the OSIRIS framework\cite{osiris}. Parameters similar to the experiment conditions were used, with a Gaussian plasma density profile to enable trapping of electrons in the density down ramp (see ref.~\citen{He_NJP_2013} and Methods for details on the simulation). 

\begin{figure}
\centering
\includegraphics[width=160 mm]{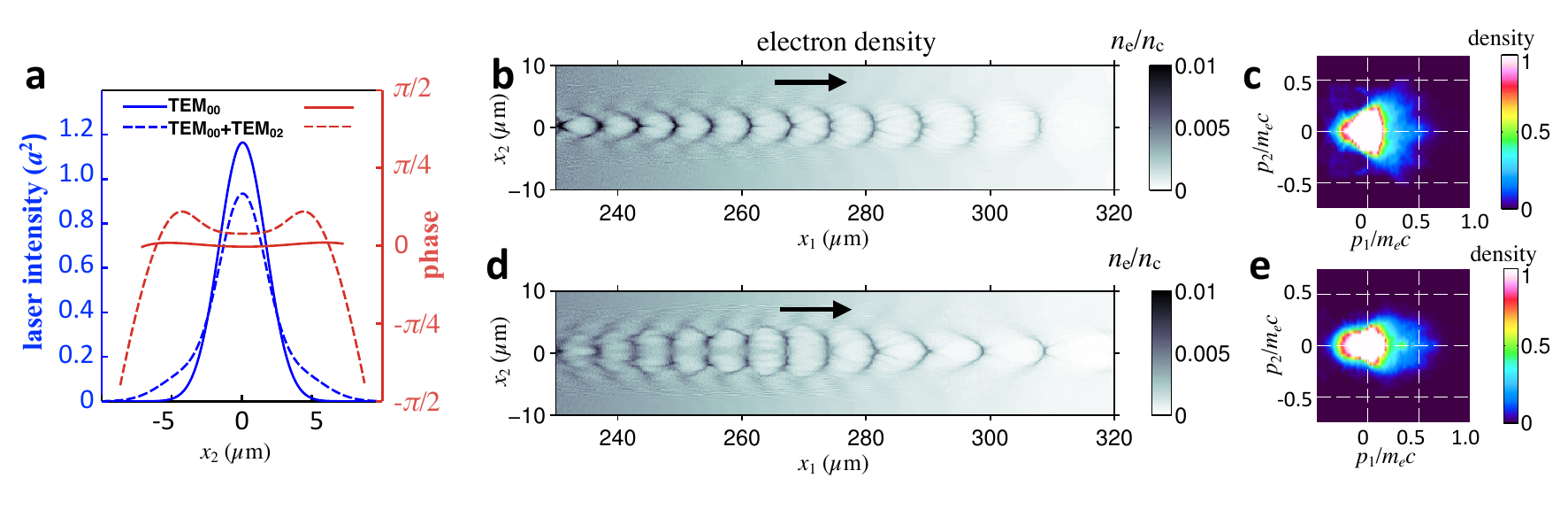}
\caption{\textbf{Modelling the effect of laser phase front using coherently superimposed modes.} (a) Transverse laser intensity profile and phase at the focal plane ($x_1=215$ $\mu$m) for a single Gaussian mode (solid curves) and a superimposed mode (dashed curves) obtained in the 2D simulations. Plotted are the laser pulses at  vacuum focus. (b)(d) are snapshots showing the different plasma wave structures around the time ($t=1.28$ ps) and location of trapping, driven by a laser pulse of single mode (b) and superimposed mode (d). The laser pulse propagates to the right. The phase space $p_2\mbox{-}p_1$ distribution of the accelerated electrons shown for (c) single mode and (e) superimposed mode before the electrons exit the simulation box ($t=2.25$ ps).}
\label{sim}
\end{figure}

It was previously shown in ref.~\citen{Cor_PRSTAB_2011}  that the focusing fields of laser plasma accelerators can be controlled by tailoring the transverse intensity profile of the laser pulse using higher-order modes, where generalization to 3D was also discussed. Here, we simulated a laser pulse with a fundamental Gaussian mode (TEM$_{00}$) or a coherent superposition of a fundamental (TEM$_{00}$) and a second-order Hermite-Gaussian (TEM$_{02}$) mode (Fig.~\ref{sim}a). Although the plasma wave has a larger amplitude when it is driven by a single mode laser pulse, the wake phase front evolves a backward curvature when electrons are trapped and accelerated (see top panel in Supplementary Movie 1 and Fig.~\ref{sim}b). Contrastingly, the evolution of the wakefield driven by the laser pulse with additional mode forms a flatter plasma phase front \emph{at the point of trapping} (Fig.~\ref{sim}d).

In Fig.~\ref{sim}c and e, the momentum distribution of the forward accelerated electrons has a larger transverse spread for the single mode laser pulse compared to the one with the addition of higher order modes. This is a consequence of the different trapping conditions and accelerating fields from the coherent plasma wakefield structure, which is governed by the structure of the driving laser pulse. In a comparative test to show this effect is not simply due to a lower intensity, we repeated the simulation using a single fundamental Gaussian mode laser pulse with a larger focal spot with the same peak intensity as that with the superimposed modes. The wakefield evolution shows very similar response as Fig.~\ref{sim}b and does not develop a flatter phase front as seen in Fig.~\ref{sim}d. The subsequently accelerated electrons have very similar divergence to that in Fig.~\ref{sim}c, eliminating the possibility that the improvement comes from either a high intensity effect or a simple change in $f$-number. \emph{Note that we are not saying that this mixed-Hermite-Gaussian mode is the optimal pulse in the experiment; this is simply an illustration of how small changes in pulse shape can have significant effects on electron beam properties.}

\section{Discussion}
When a particular wavefront of laser light interacts with plasma, it can affect the plasma wave structures and trapping conditions of the electrons in a complex way. For example, Raman forward scattering, envelope self-modulation, relativistic self-focusing, and relativistic self-phase modulation\cite{Mori_JQE_1997} and many other nonlinear interactions modify both the pulse envelope and phase as the pulse propagates, in a way that cannot be easily predicted and that subsequently dictates the formation of plasma waves. Moreover, under realistic experimental conditions, ionization dynamics before the laser pulse reaches the vacuum focus can also modify the phase of the driving pulse. Ideally, the light interacts in such a way as to generate large amplitude plasma waves with electric field structures that accelerate electrons with small divergence, high charge etc. Because of the complicated interaction, it is difficult to determine a laser phase profile that will lead to such a plasma structure. However, such unforeseeable conditions were successfully revealed by \emph{using the evolutionary algorithm method}, with the result that the electron charge can be increased and emitted in a very well collimated beam.

Here we have implemented coherent control of a nonlinear plasma wave and demonstrated an order of magnitude improvement in the electron beam parameters. The laser beam optimized to generate the best electron beam was not the one with the `best' focal spot. Control and shaping of the electron energy distribution was observed to be less effective, but was still possible. The capability for wavefront control was also limited by the number of actuators and maximum deformation of the deformable mirror used in our experiments. In addition, this work was performed using adaptive optics, but it is clear that coherent control of plasma waves should be possible in a variety of configurations, for example by using an acousto-optic modulator to control the temporal phase of the driving pulse. Recently developed techniques\cite{Li_PRL_2014,Savert_2014} for single-shot diagnosis of plasma wave structures may provide an avenue for direct control of the plasma evolution.

The concept of coherent control for plasmas opens new possibilities for future laser-based accelerators. Although still at the stage of fundamental research, laser wakefield accelerators are showing significant promise. In principle, such improvements could be integrated into next generation high-power laser projects, such as ICAN\cite{ICAN}, based on coherent combination of many independent fibers, taking advantage of both their high repetition rate and controllability.
The stability and response of the wakefield to laser conditions, such as phase front errors, is not well understood, but is crucial for the success of laser wakefield acceleration as a source of relativistic electrons and secondary radiation. For example, the presence of an asymmetric laser pulse was shown to affect the betatron oscillations and properties of x-rays produced in laser wakefield accelerators\cite{Glinec_EPL_08,Mangles_APL_2009,Popp_PRL_2010,Schnell_NatComm_2013}. Implementing the methods of this study should enable a significantly improved understanding and control of the wakefield acceleration process with regard to stability, dark current reduction and beam emittance.

\begin{figure}
\centering
\includegraphics[width=85 mm]{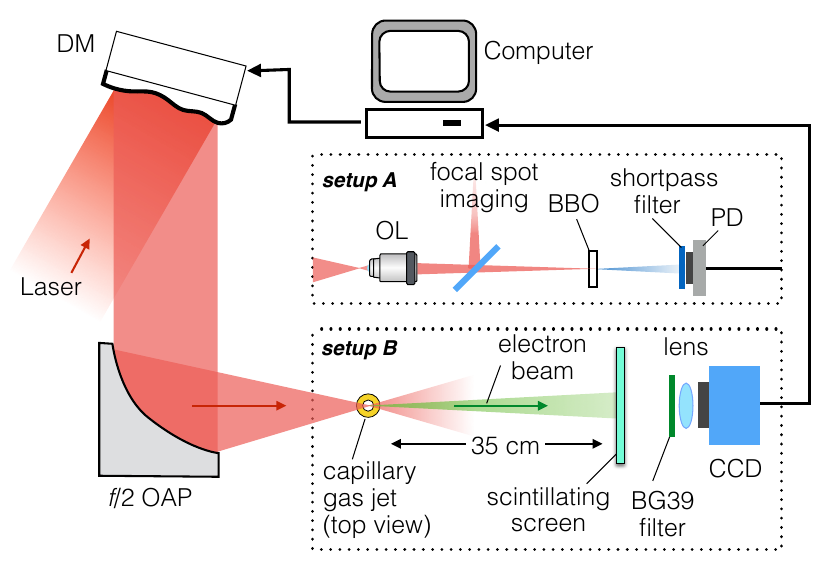}
\caption{\textbf{Experimental set-up.} DM: deformable mirror; OAP: off-axis parabolic mirror, 50 mm diameter; OL: objective lens; PD: photodiode. A: conventional focal spot optimization using second-harmonic generation; B: setup for direct optimization of the electron signal from the laser plasma accelerator.}
\label{setup}
\end{figure}

\begin{figure}
\centering
\includegraphics[width=85 mm]{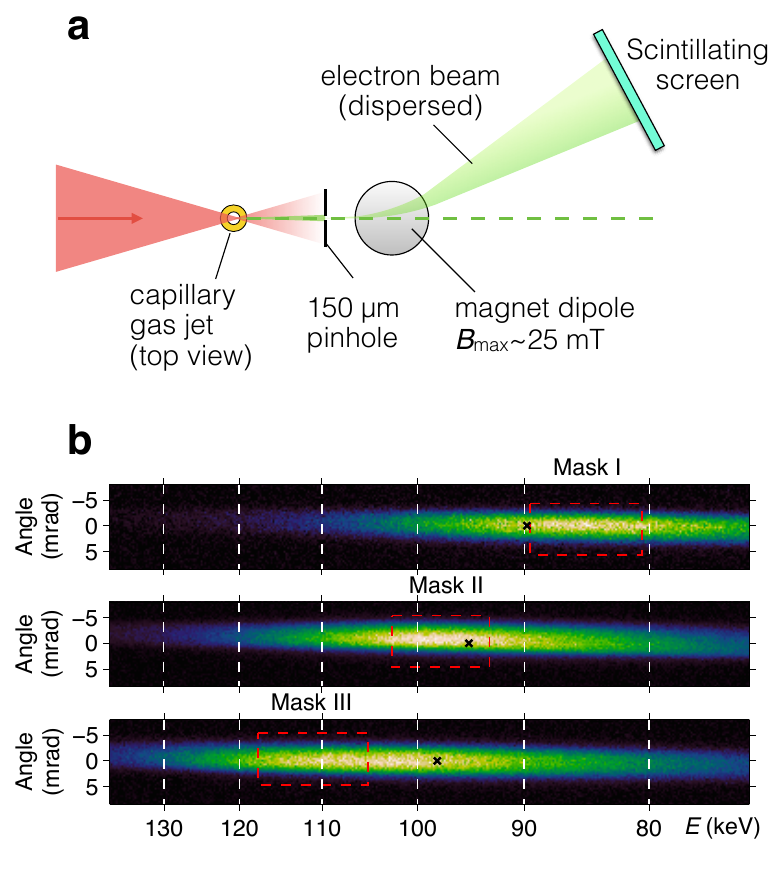}
\caption{\textbf{Control of electron energy distributiion.} (a) Schematic set-up for measuring electron energy distribution using dipole magnets. (b) Raw data showing the dispersed electron signal after genetic algorithm optimization using three different image masks. The location of the mask is indicated by the red rectangle and the black cross ($\times$) represents the final mean energy for each spectrum. 50 shots were integrated for each spectrum.}
\label{espec}
\end{figure}

\begin{methods}
\subsection{Laser System}
The Relativistic Lambda Cubed laser ($\lambda^3$) produces 30 fs pulses of 800 nm light at a repetition rate of 500 Hz with an ASE (Amplified-Spontaneous-Emission) intensity contrast of $\sim10^8$ around 1 ns before the main pulse. The system is seeded by a Femto-Laser Ti:sapphire oscillator, which generates 12 fs pulses and has a companion carrier envelope phase locking system. An RF addressable acousto-optic filter called a Dazzler controls the spectral amplitude and phase of these pulses. Selected pulses from the Dazzler train are stretched to 220 ps in a low-aberration stretcher and amplified to 7 mJ in a cryogenically cooled large-mode regenerative amplifier (Regen). The energy dumped from the Regen cavity is `cleaned' in a Pockels cell and used to seed a 3-pass amplifier as an upgrade from the laser system described in ref.~\citen{Hou:08}, which delivers up to 28 mJ pulses before compression. Following 71\% efficient compression, 20 mJ pulses are trimmed to 18 mJ at the perimeter of a 47 mm-diameter, 37-actuator deformable mirror. Throughout the system, pump light is provided by a variety of internally doubled Nd-doped YAG, YLF and vanadate lasers. The output beam with its controllable wavefront is then delivered to one of five experimental areas for the production of x-rays, electron beams, ion beams, THz radiation, high-order harmonics, or warm-dense matter. 

\subsection{Electron Acceleration and Detection}
The focused laser pulse drives plasma waves by interacting with an argon gas jet flowing continuously from a 100 $\mu$m inner diameter fused silica capillary. Typically the laser axis is 300 $\mu$m above the orifice of the tubing. The laser pulses were focused by an $f/2$ off-axis parabolic mirror to a spot size of 2.5 $\mu$m FWHM with a maximum of 10 mJ energy on target. The plasma electron density is measured to be in the range (0.5-2)$\times 10^{19}$ cm$^{-3}$ using transverse interferometry.  Electrons are accelerated in the density down ramp with a final energy in the 100 keV range\cite{He_NJP_2013} and detected by a high resolution scintillating screen (J6677 FOS by Hamamatsu), which is placed about 35 cm downstream from the source and imaged with a lens coupled 12-bit CCD camera for a 4$\times$4 cm effective area. The scintillator sensitivity was calibrated using an electron microscope for the energy range in the spectrum measurement. Electron beam charge was estimated using the calibrated scintillator response, manufacturer-provided information for the CCD camera (gain, quantum efficiency etc.) and the measured effective numerical aperture of the imaging system.

\subsection{Focal characterization and wavefront measurement}
The amplified laser beam was attenuated by using a half-wave plate and the polarization dependent properties of the compressor grating of the laser system. \O 25 mm neutral density filters (Thorlabs, Inc.) were inserted in the exit beam after the compressor and before a telescope beam expander. The laser focus was imaged by a 60$\times$ microscope objective lens (Newport Corporation, M-60X) onto a 8-bit CCD camera for focal characterization (cf setup A in Fig.~\ref{setup}). A Shack-Hartmann wavefront sensor (Flexible Optical BV) was used to determine the relative wavefront change between different deformable mirror configurations. The sensor, which consists of a $30\times 30$ microlens array having a focal length of 3.5~mm and 150~$\mu$m pitch, was directly placed in the path of the converging beam after the focusing parabolic mirror. Reconstruction of the wavefront was performed using the FrontSurfer analysis software (Flexible Optical BV), typically with  $\approx450$ measured local wavefront slopes and RMS error on the order of 0.05$\lambda$. Rotating the neutral density filters and the half-wave plate did not change the focal spot or the wavefront measurement significantly, insuring the wavefront distortion introduced by attenuation was negligible.

\subsection{Deformable Mirror and Genetic Algorithm}
The deformable mirror (AOA Xinetics) has a 47-mm clear aperture of a continuous face sheet with 37 piezoelectric actuators arranged on a square grid spaced 7 mm apart. The maximum stroke used in this experiment is about 2 $\mu$m. 

The mirror shape is controlled by a genetic algorithm, which is a method mimicking the process of natural selection and routinely used to generate optimal solutions in complex systems with a large number of variables. The \emph{genetic representation} in our experiments comprises a set of 37 independent voltage values for the deformable mirror actuator array. A \emph{fitness function} is designed to produce a single \emph{figure of merit} (FOM) to evaluate how close the solution is to the goal.

In the electron beam profile optimization experiment, FOM is computed as follows:
\begin{equation}
FOM=\sum_{\substack{{(i,j)}\\\mathbf{r_{ij}}\neq \mathbf{r_0}}}\frac{I_{ij}}{|\mathbf{r_{ij}}-\mathbf{r_0}|^n}
\label{FOM2}
\end{equation}
where $I_{ij}$ is the pixel intensity for every pixel $(i,j)$ in the whole image and $\mathbf{r_0}$ is a coordinate point in the image used as an optimization target. The power factor $n>0$ gives higher weighting to those pixels closer to the target (inverse distance weighting).

In the experiment to control the energy spectrum, FOM is calculated using the following formula given a pre-defined image mask,
\begin{equation}
{\rm FOM}=\left(1-\frac{{\rm mean\; intensity\;}{\rm outside\; mask}}{{\rm mean\;intensity\;}{\rm of \;whole\; image}}\right)
\times {\rm mean\; intensity\;inside\;mask}
\label{FOM1}
\end{equation}
The mean intensity is the sum of the pixel counts divided by the number of pixels for a defined region. A rectangular mask was used in the experiment as specified by the region enclosed by the red dashed lines in Fig.~\ref{espec}b.

\subsection{Numerical simulations}
The 2D PIC simulations were performed in a stationary box of the dimensions $573\times 102$ $\mu$m with $10000\times 600$ cells and 4$\times$4 particles-per-cell. A Gaussian plasma density profile was used in the propagation dimension ($x_1$), peaked at $x_1=200~\mu$m with a full-width-at-half-maximum of 120 $\mu$m and a maximum electron density of 0.005$n_c$, where $n_c$ is the plasma critical density. The laser pulse was initialized at the left edge of the simulation window and focused at 215 $\mu$m in the density down ramp. In 2D geometry, the transverse intensity profile of the laser pulse for fundamental Gaussian mode (TEM$_{00}$) has the form $a_0^2\exp(-2x_2^2/w_0^2)$, and the second-order Hermite-Gaussian mode $a_2^2\exp(-2x_2^2/w_2^2)(8x_2^2/w_2^2-2)^2$, where $a_{0,2}$ is the normalized vector potential and $w_{0,2}$ is the beam waist parameter. The two modes are coherently superimposed in the same plane of polarization. Here we used even-order Hermite-Gaussian mode (TEM$_{02}$) for its symmetric property. A phase difference of $\pi/8$ was applied at the beam waist between the two modes to simulate variations in the optical phase front condition. The beam waist was positioned to account for focal shift as a result of coherent superposition of two modes such that the location of the maximum on-axis laser intensity was the same (at $x_1=215~\mu$m) for all simulation runs. The simulation parameters are $a_0=1.08$ and $w_0=3.31$ $\mu$m for the Gaussian mode alone, or $a_0=1.0$, $a_2=0.15$ and $w_0=w_2=3.31$ $\mu$m for the superimposed mode. 
\end{methods}

\begin{addendum}
\item This work was supported by DARPA (Contract No.  N66001-11-1-4208), the NSF (Grant No. 0935197), NSF CAREER (Grant No 1054164), the AFOSR Young Investigator Program (Grant No. FA9550-12-1-0310) and MCubed at the University of Michigan. The authors acknowledge the OSIRIS Consortium, consisting of University of California, Los Angeles (UCLA) and Instituto Superior T\'ecnico (IST) (Lisbon, Portugal), for the use of the \textsc{osiris} 2.0 framework. This research was supported in part through computational resources and services provided by Advanced Research Computing at the University of Michigan, Ann Arbor.

\item [Author contributions] 
Z.H, B.H., J.A.N. designed and carried out the experiment. Z.H. performed all data analysis and the simulations. V.L. and B.H. developed the LabView program. K.M.K and A.G.R.T. directed and guided the project. Z.H. and A.G.R.T. wrote the paper. All authors discussed the results and contributed to the manuscript.

\item[Competing Interests] The authors declare that they have no competing financial interests.
\item[Correspondence] Correspondence and requests for materials should be addressed to Z.-H. He. (email: zhhe@umich.edu).
\end{addendum}

\end{document}